**Short title for running heads**: Law of brevity in macaque communication

# The Law Of Brevity In Macaque Vocal Communication Is Not An Artifact Of Analyzing Mean Call Durations

**Stuart Semple[1], Minna J. Hsu[2], Govindasamy Agoramoorthy[3] & Ramon Ferrer-i-Cancho[4*]**

[1]Centre for Research in Evolutionary Anthropology, University of Roehampton, UK

[2]Department of Biological Sciences, National Sun Yat-Sen University, Taiwan, Republic of China

[3]College of Environmental and Health Sciences, Tajen University, Taiwan, Republic of China

[4]Complexity and Quantitative Linguistics Lab, Universitat Politècnica de Catalunya, Barcelona (Catalonia), Spain

*Author for correspondence:

Ramon Ferrer-i-Cancho

Phone: +34 934137870

Fax: +34 934137787

Email: rferrericancho@lsi.upc.edu

Postal address: Complexity and Quantitative Linguistics Lab, Departament de Llenguatges i Sistemes Informàtics, TALP Research Center, Universitat Politècnica de Catalunya, Campus Nord, Edifici Omega, Jordi Girona Salgado 1-3, 08034 Barcelona (Catalonia), Spain

# Abstract

Words follow the law of brevity, i.e. more frequent words tend to be shorter. From a statistical point of view, this qualitative definition of the law states that word length and word frequency are negatively correlated. Here the recent finding of patterning consistent with the law of brevity in Formosan macaque vocal communication (Semple et al., 2010) is revisited. It is shown that the negative correlation between mean duration and frequency of use in the vocalizations of Formosan macaques is not an artifact of the use of a mean duration for each call type instead of the customary 'word' length of studies of the law in human language. The key point demonstrated is that the total duration of calls of a particular type increases with the number of calls of that type. The finding of the law of brevity in the vocalizations of these macaques therefore defies a trivial explanation.

# Introduction

In his pioneering work, G. K. Zipf observed that more frequent words tend to be shorter, and attributed this phenomenon to a general principle of least effort (Zipf, 1949). At the level of the dependency between length and frequency, that principle can be considered an informal precursor of a compression principle, i.e. assigning smaller lengths to more frequently used words, that has been studied with mathematical rigor in information theory (Cover & Thomas, 2006). The law of brevity has been reported for many languages (e.g., Zipf, 1949; Straus et al., 2007; Jayaram & Vidya, 2009). Recently, parallels of the law of brevity have been investigated in the behavior of other species. A negative correlation between frequency and size (in behavioural units) of surface behavioral patterns has been reported for dolphins (Ferrer-i-Cancho & Lusseau, 2009), and a negative correlation has been found between frequency of use and duration of vocalizations of Formosan macaques (Semple et al., 2010). However, support for this law has not been found in analyses of the vocalizations of two New World primates, common marmosets and golden-backed uakaris (Bezerra et al., 2011). The universality of the law in the behavior of other species is a matter of current discussion (Ferrer-i-Cancho & Hernández-Fernández 2012, Bezerra et al., 2011).

In the study by Semple et al. (2010), the law of brevity was studied by means of a correlation analysis of the relationship between mean call duration and frequency of use (the latter being quantified as the number of calls of each type recorded). It has been argued that a negative correlation between two variables, $X$ and another $Z=Y/X$ (e.g., $Z$ is the mean duration, $X$ is the frequency of use and Y is the total duration) is an unavoidable consequence of the definition of $Z$ as quotient involving $X$ because then $Z \sim 1/X$. (Solé, 2010). The argument is mathematically flawed, because $Z \sim 1/X$ requires that $X$ and $Y$ are uncorrelated (Hernández-Fernández et al., 2011; Li, 2012). However, the important message for quantitative linguistics researchers is that a negative correlation between $Z=Y/X$ and $X$ could potentially be a trivial consequence of the independence between $X$ and $Y$.

With respect to the data analyzed by Semple et al. (2010), let $D$ be the total duration of all calls of a particular type ($D$ is the sum of all the durations of the calls of a given type). The main objective of this article is to demonstrate that the negative correlation between frequency of use $f$ and mean call duration $d = D/f$ in Formosan macaques is not a straightforward consequence of the definition of $d$ as a mean. More formally, we want to reject $d=c/f$ (where $c$ is a constant) by showing that $D$ and $f$ are correlated. This potential problem in the study of the law of brevity by Semple et al. (2010) does not concern the analysis performed by Ferrer-i-Cancho & Lusseau (2009), where the correlation analysis focused on the dependency between size of a behavioral pattern in elementary behavioral units (not a mean size) and frequency.

For some quantitative linguistics researchers, another important problem, namely the fit of a particular function, e.g., $d \sim f^b$ where $b$ is a constant, would need to be addressed as is customary in quantitative linguistics research (e.g., Jayaram & Vidya, 2009). Firstly, it should be noted that we want to stay neutral on the issue of the most appropriate function for human language or animal behavior in general in the present article. For instance, standard information theory suggests that $d \sim \log f$ could be a more appropriate function based upon optimal coding considerations (Cover & Thomas, 2006) but as far as we know, this alternative functional dependency has not been considered. Secondly, it is our aim here to contribute to defining a statistically rigorous methodology for studying via correlation analysis the law of brevity as a tendency for more frequent elements to be shorter.

## Methods

The same dataset used by Semple et al. (2010) was reanalyzed here. Pearson's and Spearman's correlations, carried out in SPSS v. 17.0, were used to explore the relationship between $D$ and $f$. As in Semple et al. (2010), two levels of analysis were considered, i.e. the whole repertoire of vocalizations ($n=35$ call types) and also just those vocalizations produced by all age classes ($n=17$ call types), as some vocalization are not produced by all age groups (Hsu et al., 2009).

**Results**

Figure 1 shows the dependency between $D$ and $f$ in the vocalizations of Formosan macaques at the two levels of analysis. Indeed, $D$ and $f$ are highly correlated in the whole data set (Pearson's correlation test: $n=35$, $r=0.597$, $p<0.001$; Spearman's correlation test: $n=35$, $r_s=0.727$, $p<0.001$), and in the reduced data set of calls given by all age classes (Pearson's correlation test: $n=17$, $r=0.671$, $p=0.003$; Spearman's correlation test: $n=17$, $r_s=0.758$, $p<0.001$).

***Insert Figure 1 around here***

**Discussion**

It has been shown that the law of brevity documented in the vocal communication of Formosan macaques defies a trivial explanation. *A priori*, uncorrelation between $D$ and $f$ would have been surprising from a mathematical perspective. Notice that $D$ is defined as a sum of the durations of each vocalization type that is bounded below by a number greater than zero. To see this, introducing some notation is necessary. $\delta_i$ is defined as the duration of the $i$-th occurrence of vocalization type of frequency $f$, and $\delta_{min}$ is defined as the minimum duration of a vocalization type. Then it follows that

$$D = \sum_{i=1}^{f} \delta_i \geq f\delta_{min}. \qquad (1)$$

Notice that $\delta_{min} > 0$ for a vocalization to be perceived. Eq. 1 and the fact that $\delta_{min} > 0$ indicate that the knowledge about the exact value of $f$ constrains the set of possible values of $D$, *i.e.* there is a bias for the growth of $D$ as $f$ increases.. The point is whether $\delta_{min}$ is large enough so that this bias really matters from a practical point of view. However, there might be other factors that determine dependency between $D$ and $f$ beyond Eq. 1 and, as has been shown, a simple correlation test can help us to show that $D$ varies with regard to $f$..

The correlation results presented here indicate that the relationship between $D$ and $f$ is not purely linear, i.e. the relationship does not obey

$$D = af + b, \qquad (2)$$

where *a* and *b* are constants. Notice that $b = 0$ is needed by the definition of *D* (Eq. 1) so that $D = 0$ when $f = 0$. Thus, if *D* followed Eq. 2 then $d = D/f = a$ but *d* and *f* are significantly correlated (Semple et al. 2010). Here a very important problem for quantitative linguistics research and linguistic theory has been addressed: the statistical significance of statistical regularities of language. Contrary to what many have researchers have claimed (e.g., Miller & Chomsky, 1963; Suzuki et al., 2005; Solé, 2010), various statistical patterns of language are hard to explain in terms of artifacts or simplistic random processes such as the famous random typing experiment (Miller & Chomsky 1963); this applies not only to Zipf's law of brevity but also Zipf's law for word frequencies in both human language (Zipf, 1949) and dolphin whistles (McCowan et al., 1999), and Menzerath-Altmann's law in genomes (Ferrer-i-Cancho & Forns, 2009). Statistical patterns of language are accompanied by other statistical properties that invalidate the trivial explanations proposed so far, e.g. dependency between total duration and frequency in Formosan macaques (shown here), dependencies between behavioral context and whistle type or dependencies in whistle type sequences in dolphins (Ferrer-i-Cancho & McCowan, 2009, Ferrer-i-Cancho & McCowan 2012), and dependencies between chromosome number and total genome size (Hernández-Fernández et al., 2011).

Thus, it is clearly possible to distinguish between a simplistic explanation and a deeper explanation for a given statistical law of language. Not looking at other properties of the system for further checking, and the lack of statistically rigorous methods to evaluate the fit of a trivial explanation to actual data (see Ferrer-i-Cancho & Elvevåg, 2010 for the case of Zipf's law) has led, in our opinion, to wrong conclusions about the importance and the enormous potential of statistical patterns not only of human language, but also of vocal communication and other behaviour in non-human animals.

## Acknowledgements

This work was supported by University of Roehampton, and by the grant *Iniciació i reincorporació a la recerca* from the Universitat Politècnica de Catalunya and the grant BASMATI (TIN2011-27479-C04-03) from the Spanish Ministry of Science and Innovation.

# References

Bezerra, B. M., Souto, A. S., Radford, A. N., & Jones, G. (2011). Brevity is not always a virtue in primate communication. *Biology Letters,* **7**, 23-25. doi: 10.1098/rsbl.2010.0455

Cover, T. M., & Thomas, J. A. (2006). *Elements of information theory (2$^{nd}$ ed.)*. Hoboken, New Jersey: Wiley.

Ferrer-i-Cancho, R., & Elvevåg, B. (2010). Random texts do not exhibit the real Zipf's law-like rank distribution. *PLoS ONE, 5*, e9411. doi:10.1371/journal.pone.0009411

Ferrer-i-Cancho, R. & Hernández-Fernández, A. (2012). The failure of the law of brevity in two New World primates. Statistical caveats. *Glottotheory* 4 (1), in press. http://arxiv.org/abs/1204.3198.

Ferrer-i-Cancho, R., & Lusseau, D. (2009). Efficient coding in dolphin surface behavioral patterns. *Complexity, 14*, 23-25. doi: 10.1002/cplx.20266

Ferrer-i-Cancho, R., & McCowan, B. (2009). A law of word meaning in dolphin whistle types. *Entropy, 11*, 688-701. doi: 10.3390/e11040688

Ferrer-i-Cancho, R. & McCowan, B. (2012). The span of correlations in dolphin whistle sequences. *Journal of Statistical Mechanics*, P06002. doi: 10.1088/1742-5468/2012/06/P06002.

Ferrer-i-Cancho, R., & Forns, N. (2009). The self-organization of genomes. *Complexity, 15*, 34-36. doi: 10.1002/cplx.20296

Hernández-Fernández, A., Baixeries, J., Forns, N., & Ferrer-i-Cancho, R. (2011). Size of the whole *versus* number of parts in genomes. *Entropy, 13*, 1465-1480. doi: 10.3390/e13081465

Hsu M. J., Chen L.-M., & Agormaoorthy, G. (2005). The vocal repertoire of Formosan macaques, *Macaca cyclopis*: acoustic structure and behavioral context. *Zoological Studies, 44*, 275–294.

Jayaram, B. D., & Vidya, M. N. (2009). The relationship between word length and word frequency in Indian languages. *Glottotheory, 2*, 62-69.

Li, W. (2012). Menzerath's law at the gene-exon level in the human genome. *Complexity,* 17, 49-53.

McCowan, B., Hanser, S. F., & Doyle, L. R. (1999). Quantitative tools for comparing animal communication systems: information theory applied to bottlenose dolphin whistle repertoires. *Animal Behaviour, 57*, 409-419. doi:10.1006/anbe.1998.1000

Miller, G. A., & Chomsky, N. (1963). Finitary models of language users. In R. D. Luce, R. Bush, & E. Galanter (Eds.), *Handbook of Mathematical Psychology* (pp. 419-492). New York, NY: Wiley.

Semple, S., Hsu, M. J., & Agoramoorthy, G. (2010). Efficiency of coding in macaque vocal communication. *Biology Letters, 6*, 469-471. doi: 10.1098/rsbl.2009.1062

Solé, R. V. (2010). Genome size, self-organization and DNA's dark matter. *Complexity, 16*, 20-23. doi: 10.1002/cplx.20326

Strauss, U., Grzybek, P., & Altmann, G. (2007). Word length and word frequency. In P. Grzybek (Ed.) *Contributions to the science of text and language* (pp. 277–294). Dordrecht, The Netherlands: Springer.

Suzuki, R., Tyack, P. L., & Buck, J. R. (2005). The use of Zipf 's law in animal communication analysis. *Animal Behavior, 69*, F9–F17. doi:10.1016/j.anbehav.2004.08.004.

Zipf, G. K. (1949). *Human behaviour and the principle of least effort.* Cambridge, MA: Addison-Wesley.

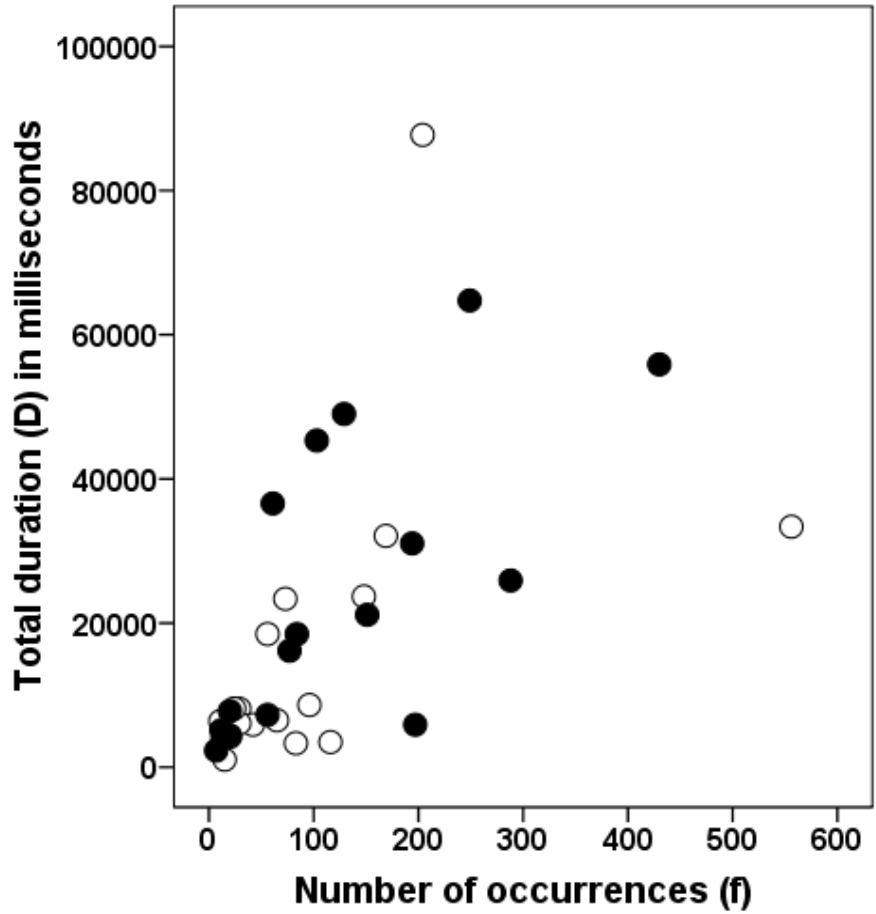

Figure 1. The dependency between total duration *D* (in milliseconds) and frequency of occurrence *f* of Formosan macaque vocalizations. Each point represents one call type. Black circles indicate calls given by members of all four age classes (adult, subadult, juvenile and infant); open circles are all other calls in the repertoire.